\documentclass[twocolumn,showpacs,preprintnumbers,amsmath,amssymb]{elsart3}
\usepackage{natbib}

\usepackage{graphicx} 
\usepackage{epsfig} 
\usepackage{dcolumn} 
\usepackage{bm}

\newcommand{\psip}{\psi(2S)}
\newcommand{\jpsi}{J/\psi}
\newcommand{\rar}{\rightarrow}
\newcommand{\pipi}{\pi^+\pi^-}
\newcommand{\kst}{K^*(892)}
\newcommand{\kstark}{K^*(892)\overline{K}+c.c.}
\newcommand{\kstarkpm}{K^*(892)^+K^-+c.c.}
\newcommand{\kstarknn}{K^*(892)^0\overline{K}^0+c.c.}
\newcommand{\ks}{K_S^0}

\def\Journal#1&#2&#3(#4){#1{\bf #2}, #3 (#4)}

\def\NPB{Nucl.  Phys.  {\bf B }}
\def\PLB{Phys.  Lett.  {\bf B }}
\def\PRL{Phys.  Rev.  Lett.  }
\def\PRD{Phys.  Rev.  {\bf D }}

\def\etal{{\it et al.}}

\def\bec{\begin{center}}
\def\eec{\end{center}}

\begin{document}
\begin{frontmatter}
\title{ Observation of the decay $\psip\rar\kstark$\\ }

\date{\today}

\maketitle

\begin{center} 
M.~Ablikim$^{1}$, J.~Z.~Bai$^{1}$, Y.~Ban$^{10}$, 
J.~G.~Bian$^{1}$, X.~Cai$^{1}$, J.~F.~Chang$^{1}$, 
H.~F.~Chen$^{16}$, H.~S.~Chen$^{1}$, H.~X.~Chen$^{1}$,
J.~C.~Chen$^{1}$
% ...
\\(BES Collaboration)\\ 
\vspace{0.2cm}
$^1$ Institute of High Energy Physics, Beijing 100039, People's Republic of China\\
% ...
$^{10}$ Peking University, Beijing 100871, People's Republic of China\\
% ...
$^{16}$ University of Science and Technology of China, Hefei 230026, People's Republic of China\\
% ...
\end{center}

\begin{abstract}
Using 14 million $\psi(2S)$ events collected with the BESII detector,
branching fractions of $\psip\rar\kstarkpm$ and $\kstarknn$ are
determined to be: $B(\psip\rar\kstarkpm)=(2.9^{+1.3}_{-1.7}\pm0.4)\times
10^{-5}$ and $B(\psip\rar\kstarknn)=(13.3^{+2.4}_{-2.7}\pm1.7)\times
10^{-5}$. The results confirm the violation of the ``12\% rule'' for
these two decay channels. A large isospin
violation between the charged and neutral modes is observed.
\end{abstract}

\begin{keyword}
\PACS 13.25.Gv \sep 12.38.Qk \sep 14.40.Gx
\end{keyword}

\end{frontmatter}

\section{Introduction}
One of the longstanding mysteries in heavy quarkonium physics is the
strong suppression of $\psip$ decays to the vector-pseudoscalar (VP)
meson final states, $\rho\pi$ and $\kstark$, referred to as
the ``$\rho\pi$ puzzle''.  In perturbative QCD (pQCD), hadronic decays of
the $\jpsi$ and $\psip$ are expected to proceed dominantly via three
gluons or a single direct photon, with widths proportional to the
square of the $c\bar{c}$ wave function at the origin, which is well
determined from dilepton decays. Thus, it is reasonable to expect, 
for any hadronic final state h, the $\jpsi$ and $\psip$
decay branching fractions should satisfy the so-called 
``12\% rule''\cite{qcd15}
\begin{eqnarray*}
Q_h= \frac{B(\psip\rar h)}{B(\jpsi\rar h)}
   \simeq\frac{B(\psip\rar e^+e^-)}{B(\jpsi\rar e^+e^-)}\simeq 12\%,
\end{eqnarray*}
where the leptonic branching fractions are taken from the Particle
Data Group (PDG) tables \cite{PDG}.
It was first observed  by the MarkII
experiment that, while this rule works reasonably well for a
number of exclusive hadronic decay channels,
 it is severely violated
for the vector-pseudoscalar meson (VP) final states, $\rho\pi$ and
$\kstarkpm$~\cite{rhopi}. Preliminary BESI results confirm the MarkII measurements at
higher sensitivity~\cite{ichep97}. This anomaly has generated much
interest and led to a number of theoretical
explanations~\cite{qcd15_theory, suzuki}. More precise experimental results
are required to distinguish between them. 

In this paper, the branching fractions of charged and
neutral $\psip\rar \kstark$ decays are reported, based on a sample of
$14.0\times10^6 (1\pm4\%)$ $\psi(2S)$ events~\cite{N_psip}
collected with the Beijing Spectrometer (BESII)~\cite{BES-II} at the Beijing
Electron-Positron Collider (BEPC).

   In this analysis, a GEANT3
based Monte Carlo package (SIMBES) with detailed
consideration of the detector performance (such as dead
electronic channels) is used.
The consistency between data and Monte Carlo has been carefully checked in
many high purity physics channels, and
the agreement is reasonable~\cite{J3pi}.
The generator  KSTARK~\cite{kstark}, which simulates
$\psip\rar\kstark$ events, together with SIMBES, is used
to determine detection efficiencies.

\section{Event selection}
Candidate events for this decay mode have the final state
$\ks K^\pm\pi^\mp\rar\pi^+\pi^-K^\pm\pi^\mp$.
They are required to
satisfy the following general selection criteria:
(i) The number of charged particles must be equal to four with net charge
     zero.
(ii)  Each charged track is required to be well fitted
to a three dimensional helix and be in the polar angle
region $|\cos\theta|<0.8$. 
%
(iii)  Background from $\psi(2S)\rightarrow\pi^{+}\pi^{-}J/\psi$, 
$J/\psi\rightarrow X$ is removed by 
requiring that the recoil mass of any two oppositely charged tracks satisfies
\begin{eqnarray*}
m_{recoil}^{\pi\pi}
 &= & \sqrt{(E_{cm}-E_{+}-E_{-})^{2}-(\vec{p}_{+}+\vec{p}_{-})^2}  \\
 &\not\in& (3.05,3.15) \hbox{ GeV/$c^2$},
\end{eqnarray*}
where $E_+ (E_-)$ and $\vec{p}_+ (\vec{p}_-)$ are the assumed $\pi^+(\pi^-)$ energy and momentum, respectively.

\begin{figure}[h] \centering
\includegraphics[height=5cm]{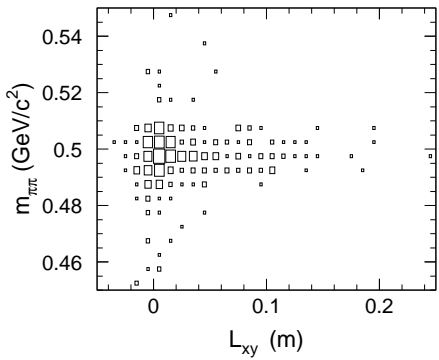}
\caption{ \label{kshort} The distribution of $m_{\pi\pi}$ versus $L_{xy}$ for
   oppositely charged track pairs in $\pi^+\pi^-K^\pm\pi^\mp$ candidate events.}
\end{figure}

The $\ks$ is identified through the decay $\ks\rar\pipi$. The
intersections of all pairs of oppositely charged tracks, assumed to be
pions, are found as described in Ref.~\cite{kskl}. The intersection is
regarded as the secondary vertex.  Figure~\ref{kshort} shows the
scatter plot of the $\pipi$ invariant mass versus the decay length in
the transverse plane ($L_{xy}$) for candidate events.  The cluster of
events with mass consistent with the nominal $\ks$ mass indicates a
clear $\ks$ signal.  Requirements $|m_{\pi^+\pi^-}-0.497|<0.018$ GeV/$c^2$
and $L_{xy}>0.01$ m are used to remove background from non-$\ks$ events.
%

Events are kinematically fitted with four constraints (4C) to the hypothesis 
$\psip\rar\pi^+_1\pi^-_2 K^\pm\pi^\mp$. Here $\pi^+_1\pi^-_2$ are
associated with the $\ks$ decay, as determined above.
For the remaining two tracks, the identification as $K$ or $\pi$ is done
in the following way: (i) If the momentum of one track is larger than 
1.34 GeV/$c$, that track is assigned to be a kaon and the other track a pion.
(ii) If the momenta are both less than 1.34 GeV/$c$, the fit is
applied to the two possible combinations, and the one with the smaller $\chi^2$
is chosen.
The confidence level of the selected 4C fit is required to be larger than 0.01.

In addition, the combined chisquare
($\chi^2_{com}$) for the assignment $\psi(2S)\rightarrow
\pi^+\pi^-K^\pm\pi^\mp$ is required to be smaller than those for the 
alternative  hypotheses
$\psi(2S)\rightarrow \pi^+\pi^-K^+K^-$ and $\psi(2S)\rightarrow
2(\pi^+\pi^-)$.
Here, the combined chisquare, $\chi_{com}^{2}$, is defined as the sum
of the $\chi^2$ values of the kinematic fit ($\chi^{2}_{kine}$) and those from
the particle identification assignments of the four tracks 
($\chi^{2}_{PID}$)~\cite{BES-VT}: 
$\chi_{com}^{2}=\sum_{i}\chi^{2}_{PID}(i)+\chi^{2}_{kine}.$

\begin{figure}[h]
\begin{center}
\includegraphics[height=5.5cm, width=0.4\textwidth]{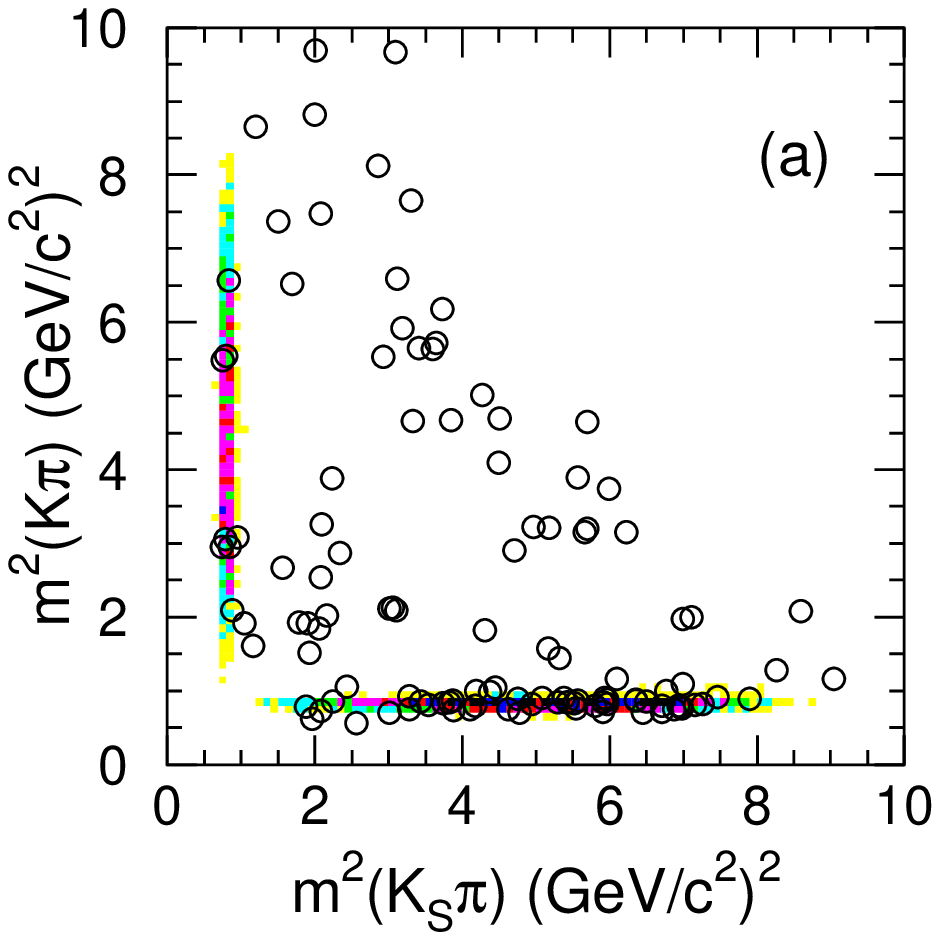}
\includegraphics[height=7cm, width=0.38\textwidth]{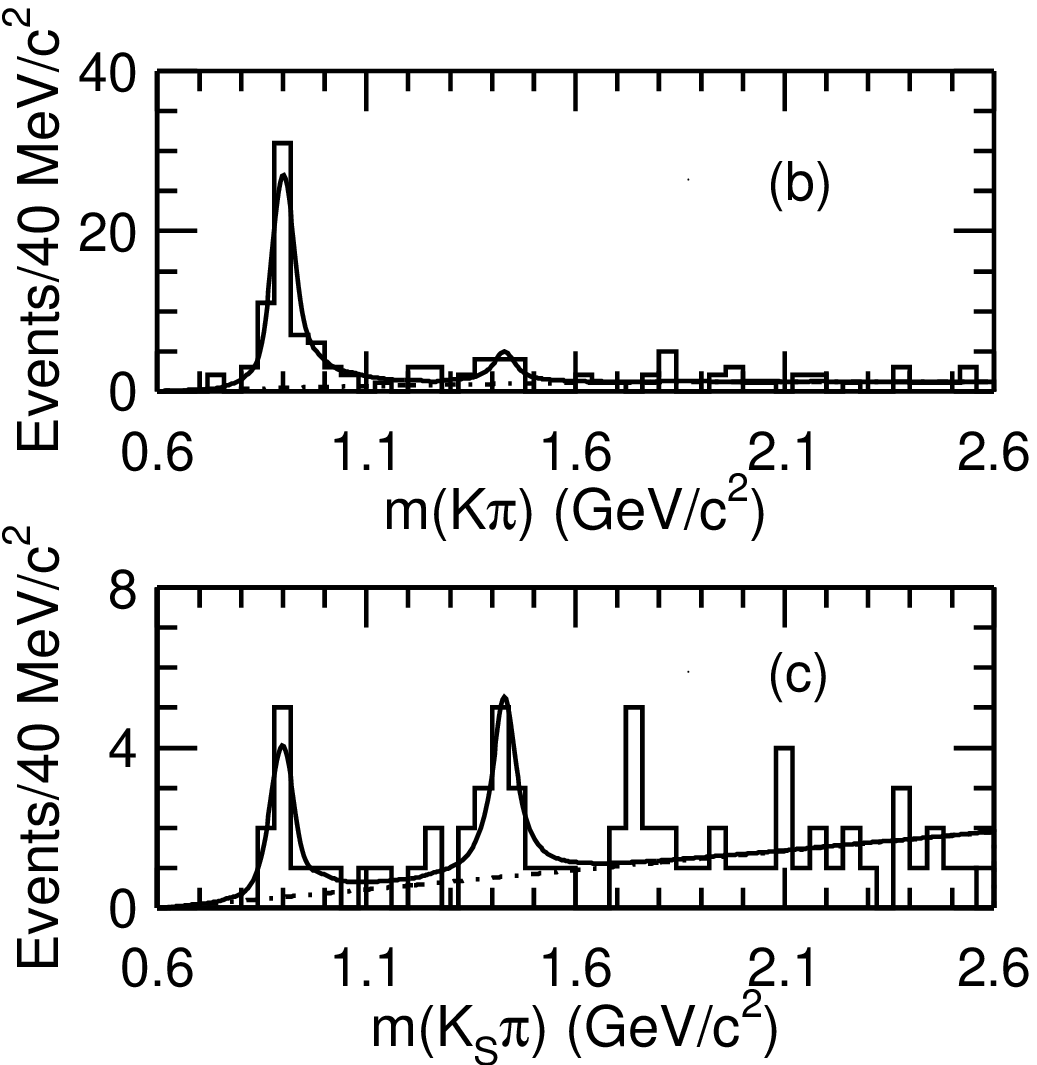}
\caption{\label{k3pi_Dalitz}
  (a) Dalitz plot, (b) $K^{\pm}\pi^{\mp}$ invariant mass, and (c)
  $\ks\pi^{\pm}$ invariant mass for $K_S^0 K^\pm\pi^\mp$ candidate
  events after the selection described in the text. The circles are data, and
  the shaded regions are simulated $\psip\rar\kstark$ events
  (the horizontal band for $\kstarknn$ and the vertical band for
  $\kstarkpm$).  The curves in (b) and (c) show the  fit described in the text.
}
\end{center}
\end{figure}

After the above selection, the Dalitz plot for $\ks K^{\pm}\pi^{\mp}$ candidate
events, shown in Fig.~\ref{k3pi_Dalitz}(a), is obtained.  Monte Carlo
simulated $\psip\rar\kstark$ events,  shown in
grey-scale, lie along a horizontal band for
$\kstarknn$ events and a vertical band for $\kstarkpm$ events.
Figure~\ref{k3pi_Dalitz}(b) shows the $K^{\pm}\pi^{\mp}$ invariant
mass after an additional requirement $m_{\ks\pi^{\pm}}>1.0$ GeV/$c^2$ to
remove $\kstarkpm$ events, and Fig.~\ref{k3pi_Dalitz}(c) shows the
$\ks\pi^{\pm}$ invariant mass after an additional requirement
$m_{K^{\pm}\pi^{\mp}}>1.0$ GeV/$c^2$ to remove $\kstarknn$ events.

Contamination from background channels, which pass the
selection criteria for the $K_S^0 K^{\pm}\pi^{\mp}$ events, 
mainly come from $\psip\rar{\gamma\chi_{cJ}}$, $\chi_{cJ}\rar$$\ks
K^+\pi^- +c.c.$, which are assumed to take place via the intermediate
state $\kstark$ Using the branching fractions taken from the
PDG~\cite{PDG}, the contamination from these channels is estimated to
be less than 0.6 events for both the charged and neutral modes.  The
events along the third side of the Dalitz plot (diagonal side) in
Fig.~\ref{k3pi_Dalitz}(a) may be due to the process
$\psip\rar\rho(2150)^{\pm}\pi^{\mp}$, $\rho(2150)^{\pm}\rar\ks
K^{\pm}$. These contaminations are not corrected for, but are included
in the systematic error, they and contribute an additional 5.6\%
systematic error for both the charged and neutral modes. The
contributions from other backgrounds are negligible.

The invariant mass spectra for $K^{\pm}\pi^{\mp}$ and $\ks\pi^{\mp}$
are fitted using the $\kst$ signal shape determined with  MC
simulation plus a second order polynomial background and an additional Breit-Wigner
function for the $K^*_J(1430)$ (described below), 
as shown in Figs.~\ref{k3pi_Dalitz}(b) and (c);  
$65.6\pm9.0$ $\kstarknn$ and $9.6\pm4.2$ $\kstarkpm$ events 
are observed. Their detection efficiencies are $(9.68\pm0.07)\%$ and $(7.25\pm0.07)\%$,
and their statistical significances are $11\sigma$ and $3.5\sigma$, respectively.

In addition, $10.5\pm5.1$ events and $11.2\pm5.3$ events near 1430 MeV
are found in the invariant mass spectra of $K^\pm\pi^\mp$ and
$K_S^0\pi^\pm$, respectively, by fitting with Breit-Wigner functions
with the means fixed at 1.43 GeV/$c^2$.  Their fitted widths are
roughly 46 MeV/$c^2$ and 100 MeV/$c^2$, and corresponding statistical
significances are $3.4\sigma$ and $3.1\sigma$, respectively.  These
events might be associated with the  $K^*_0(1430)$, $K^*(1410)$ or $K^*_2(1430)$, but
the limited statistics does not allow a determination of the spin J
(=0, 1, or 2).
Their detection efficiencies are $(9.2\pm1.0)\%$ and $(7.7\pm0.9)\%$,
where the errors reflect the unknown spin.

\section{Systematic errors and contributions from continuum}
The branching fraction is calculated from
$$B =\frac{n^{obs} } {N_{\psi(2S)}\cdot B_{int} \cdot \epsilon
  \cdot f_c},$$ where $n^{obs}$ is the number of observed $\kstark$
  events, $\epsilon$ is the detection efficiency obtained from the MC
  simulation, $f_c=(96.3\pm 3.3)\%$ is an efficiency correction factor
  from the $K_S$ reconstruction~\cite{kskl}, $N_{\psip}$ is the total
  number of $\psip$ events, and $B_{int}=1/3$ is taken as the branching
  fraction for $\kstark\rar K_S K^{\pm}\pi^{\mp}$.  The
  $K_S\rar\pi^+\pi^-$ branching ratio was
  included in the Monte Carlo simulation.

Many sources of systematic error are considered.  Those
associated with the efficiency are determined by comparing $J/\psi$
and $\psi(2S)$ data with Monte Carlo simulations for very clean decay
channels, such as $\jpsi\rar\rho\pi$, $\kstark$, and $\psi(2S) \rar
\pi^+ \pi^- J/\psi$, which allow the determination of systematic
errors associated with the MDC tracking efficiency, kinematic fitting,
particle identification, photon selection efficiency and other experimental
effects~\cite{systematics}.
 The uncertainties of the background shapes and the total number of
$\psi(2S)$ events are also sources of systematic errors.
The total systematic errors for charged and neutral $\kstark$ mode
are 14.0\% and 12.6\%, respectively. Table~\ref{SysError} summarizes 
the systematic 
errors.

\begin{figure}[hb]
\begin{center}
\includegraphics[height=4.5cm,width=0.33\textwidth]{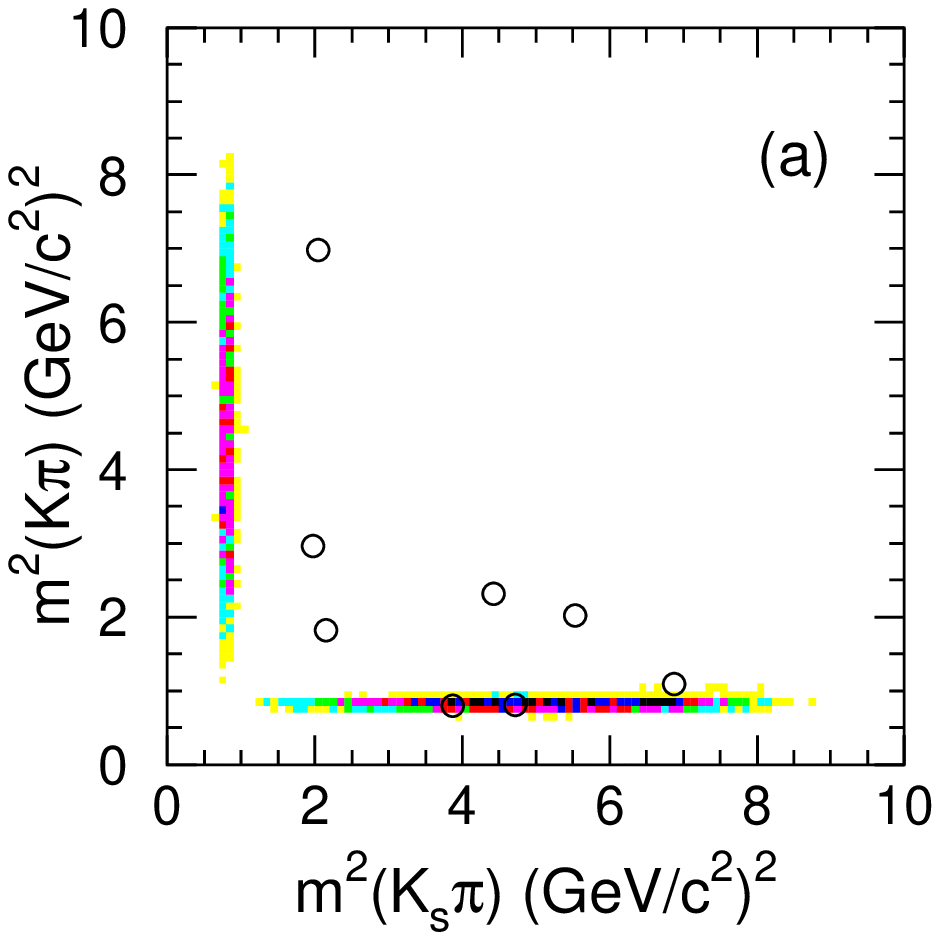}
\includegraphics[height=5.cm,width=0.32\textwidth]{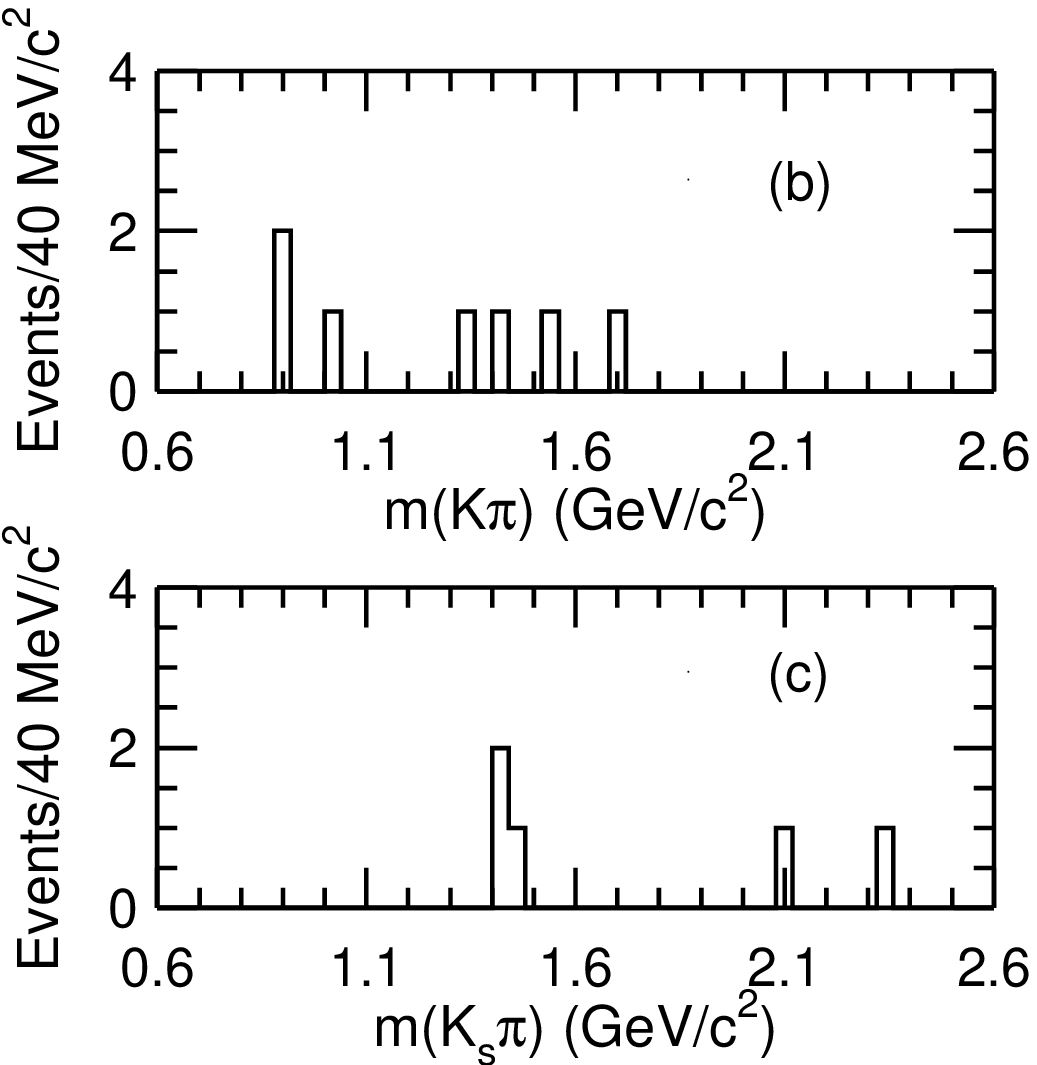}
\caption{\label{k3pi_365_fit} (a) Dalitz plot, (b) $K^{\pm}\pi^{\mp}$
invariant mass, and (c) $\ks\pi^{\pm}$ invariant mass for $K_S^0
K^\pm\pi^\mp$ candidate events in the $\sqrt{s}=3.65$ GeV data
sample. The circles are data, and the shaded regions
are simulated $\psip\rar\kstark$ events (the horizontal band for
$\kstarknn$ events and the vertical band for $\kstarkpm$ events).
}
\end{center}
\end{figure}

Contributions from the continuum $e^+ e^- \rightarrow
\gamma^*\rightarrow$ hadrons \cite{wangp,wangp2} are estimated using a 
sample taken at $\sqrt s=3.65$ GeV of $6.42\pm0.24$ pb$^{-1}$~\cite{Lum}, 
about one-third of the integrated luminosity at the
$\psi(2S)$.  In $K^*(892)^0 \overline{K}^0+c.c.$, $2.5^{+2.6}_{-1.8}$ events are
observed, as shown in Fig.~\ref{k3pi_365_fit}, while no events are
observed in the $K^*(892)^{+}K^{-}+c.c$.
channel, which corresponds to $0.0^{+1.3}_{-0.0}$
events at the 68.3\% confidence level~\cite{Jconrad}.

\begin{table}[h]
\begin{center}
\caption{\label{SysError} 
 Summary of systematic errors ($\%$).}
\begin{tabular}{|c|c|c|}  \hline \hline
  &$K^*(892)^{\pm}K^{\mp}$ &  $\kstarknn$ \\ \hline
Tracking          &8.0 &8.0 \\
kine. fit         &2.9 &2.9 \\
Bkgd shape        &5.0 &5.1 \\
Bkgd comtam.      &8.4 &5.7 \\
$\ks$ reconstion  &3.4 &3.4 \\
MC statistics     &1.1 &1.0\\ \hline
$N_{\psi(2S)}$    & \multicolumn{2}{c|}{4.0} \\  \hline
sum               & 14.0 & 12.6 \\ \hline \hline
\end{tabular}
\end{center}
\end{table}


\section{Results}

Table~\ref{BESII_VP_CS} summarizes the observed numbers of events,
detection efficiencies, and corresponding cross sections for
$\kstark$ channels at
$\sqrt{s}=$ 3.65 GeV and $m_{\psip}$, respectively. Here, detection
efficiencies at $\sqrt{s}=$ 3.65 GeV include the effect
of initial state radiation. In addition, the cross sections of 
$e^+e^-\rar K^*_J(1430)^{0}K^0+c.c.\rar\ks
K^{\pm}\pi^{\mp}$ and
$e^+e^-\rar K^*_J(1430)^{+}K^-+c.c.\rar\ks
K^{\pm}\pi^{\mp}$ at 3.686 GeV are determined to be $6.0\pm2.9\pm1.0$ pb
and $7.7\pm3.6\pm1.3$ pb, respectively, where the 
first errors are statistical and the second  systematic, 
including the uncertainty from the unknown spin. The corresponding upper
limits are 11.4 pb and 13.7 pb, respectively, at the $90\%$ confidence 
level.

\begin{table}
\begin{center}
\caption{\label{BESII_VP_CS}
 Observed cross sections for $e^+e^-\rightarrow \kstark$ at $\sqrt{s}=3.65$
    GeV and 3.686 GeV.} \vskip 2pt
\begin{tabular}{c|c|c|c|c}  \hline \hline
Channels & $\sqrt{s}$  & $N^{obs}$  & $\epsilon$ & $\sigma$  \\
         & (GeV) & & (\%) & (pb) \\ \hline
 $\kstarkpm$ & & $0.0^{+1.3}_{-0.0}$ & 5.7 & $<21$ (90\% C.L.) \\
 $\kstarknn$ &3.65 & $2.5^{+2.6}_{-1.8}$ & $7.7$ &  $16^{+16}_{-11}\pm2$  \\
             & & & & $<42$ (90\% C.L.) \\ \hline
$\kstarkpm$ & 3.686 & $9.6\pm4.2$ & 7.3 & $20.9\pm9.1\pm2.9$ \\
$\kstarknn$ & & $65.6\pm9.0$ & $9.7$ & $107\pm15\pm13$   \\  \hline \hline
\end{tabular}
\end{center}
\end{table}

\begin{table}
\begin{center}
\caption{\label{BESII_VP}
  Branching fractions measured for $\psi(2S)\rightarrow \kstark$. 
  The corresponding $J/\psi$ branching fractions~\cite{PDG} and 
   the ratios $Q_h=\frac{B(\psi(2S))}{B(J/\psi)}$ are also given. } \vskip 2pt
\begin{tabular}{c|c|c|c}  \hline \hline
Channels & $B(\psi(2S))$ & $B(J/\psi)$ & $Q_h$ \\
         & ($\times 10^{-5}$) & ($\times 10^{-4}$)  & (\%) \\ \hline
$\kstarkpm$  & $2.9^{+1.3}_{-1.7}\pm0.4$ & $50\pm4$ & $0.59^{+0.27}_{-0.36}$\\
$\kstarknn$  & $13.3^{+2.4}_{-2.8}\pm1.7$ & $42\pm4$ & $3.2\pm0.8$  \\  \hline \hline
\end{tabular}
\end{center}
\end{table}

Table~\ref{BESII_VP} lists the
branching fractions for the $\psi(2S)\rightarrow \kstark$ decay modes,
where the contributions of the continuum is subtracted incoherently
with normalized integrated luminosity without considering its interference 
with the resonant amplitude.  
The table also lists the branching fractions of $J/\psi$ decays~\cite{PDG} 
as well as the ratios of the $\psi(2S)$ to $J/\psi$ branching fractions. 
The ratio $\frac{B(\psi(2S)\rightarrow\kstarknn)}{ B(\psi(2S)\rightarrow\kstarkpm)} =4.6^{+2.9}_{-2.2}$ shows
a large isospin violation between the charged and neutral modes of
$\psip\rar\kstark$ decays.
Since the amplitudes for $\psip\rar\kstark$ decays up to first order
in the SU(3) breaking consists of two parts: the strong decay
amplitude and the electromagnetic amplitude~\cite{Jdecay}, a possible
interpretation for this large isospin violation is that the
electromagnetic decay amplitude plays an important role in
$\psip\rar\kstark$ decays, while in $\jpsi$ decays the strong decay
amplitude dominates. The results listed in Table~\ref{BESII_VP} show
that, $\psip\rar{\kstark}$ branching fractions are strongly suppressed
with respect to the pQCD expectation. The charged branching fraction
is suppressed more than the neutral one and is consistent with the
upper limit measured by the MarkII experiment ($<5.4\times10^{-5}$, at
90\% C.L.)~\cite{rhopi}. Our results marginally accommodate the predictions by Chaichian et al. and by Feldmann et al.~\cite{qcd15_theory},
 two predictions for the charged mode
branching fractions are $4.5\times 10^{-5}$ and $1.2\times 10^{-5}$
respectively, while being larger than their two predictions for the
neutral mode ($7.6\times10^{-5}$ and $5.1\times 10^{-5}$,
respectively). Ma~\cite{qcd15_theory} predicted $Q_{K^*K}$ to be
$(3.6\pm0.6)\%$, which is consistent with our measurement for the
neutral mode, but not for the charged mode.

Based on the observed cross sections in 
Table~\ref{BESII_VP_CS}, the
branching fractions of $\psip\rar \kstarkpm$ and $\kstarknn$ may be
calculated by the model proposed in Ref.~\cite{wangp2}, where all the
contributions from the continuum one-photon annihilation amplitude, the 
electromagnetic amplitude and the three-gluon amplitude of the 
$\psi(2S)$ decay, and their interferences 
are taken into account. By fitting these observed cross sections the phase 
between the electromagnetic amplitude and
three-gluon amplitude of the $\psi(2S)$ decay is constrained in the  range 
from $95^0$ to $304^0$, 
disfavors the positive solution of the orthogonal phase $\pm90^0$ 
determined from $\jpsi$ decays~\cite{suzuki}.
The branching fractions in this case are:
$ B(\psip\rar\kstarkpm)=(3.1^{+1.8}_{-1.9})\times 10^{-5}$ and
$ B(\psip\rar\kstarknn)=(13.7^{+1.8}_{-9.0})\times 10^{-5}$, where the
large errors are due to the large phase uncertainty.

 In conclusion, we present the branching 
fractions for $\psip\rar\kstarknn$ and $\kstarkpm$.
They are suppressed
with respect to the pQCD expectation, and a large isospin violation between
the charged and neutral mode is observed.
These results are compatible with those recently reported by CLEO Collaboration for the same channels in ref.~\cite{Cleo}.

   The BES collaboration thanks the staff of BEPC for their 
hard efforts. This work is supported in part by the National 
Natural Science Foundation of China under contracts 
Nos. 19991480, 10225524, 10225525, the Chinese Academy
of Sciences under contract No. KJ 95T-03, the 100 Talents 
Program of CAS under Contract Nos. U-11, U-24, U-25, and 
the Knowledge Innovation Project of CAS under Contract 
Nos. U-602, U-34(IHEP); by the National Natural Science
Foundation of China under Contract No. 10175060 (USTC); 
and by the Department of Energy under Contract 
No. DE-FG03-94ER40833 (U Hawaii).


\begin{thebibliography}{99}
\bibitem{qcd15}T. Appelquist and H. D. Politzer, {\Journal\PRL&34&43(1975)};
               A. De Rujula and S. L. Glashow, {\em ibid}, page 46.
\bibitem{PDG}Particle Data Group, S. Eidelman \etal, {\Journal\PLB&592&1(2004)}, and references therein.

\bibitem{rhopi} M. E. B. Franklin \etal, MarkII Collab., {\Journal\PRL&51&963(1983).}
\bibitem{ichep97}Y. S. Zhu (Representing BES Collab.) in Proceedings of the 28th International Conference on High Energy Physics, ed. Z. Adjuk and A. K. Wroblewski, World Scientific, 1997, p507.
\bibitem{qcd15_theory}W. S. Hou and A. Soni, {\Journal\PRL&50&569(1983)};
        G. Karl and W. Roberts, {\Journal\PLB&144&263(1984)};
        S. J. Brodsky \etal, {\Journal\PRL&59&621(1987)};
        M. Chaichian \etal, {\Journal\NPB&323&75(1989)};
        S. S. Pinsky, {\Journal\PLB&236&479(1990)};
        X. Q. Li \etal, {\Journal\PRD&55&1421(1997)};
        S. J. Brodsky and M. Karliner, {\Journal\PRL&78&4682(1997)};
        Yu-Qi Chen and Eric Braaten, {\Journal\PRL&80&5060(1998)};
        T. Feldmann and P. Kroll, {\Journal\PRD&62&074006(2000)};
        J. L. Rosner, {\Journal\PRD&64&094002(2001)};
        J. P. Ma, {\Journal\PRD&65&097506(2002)};
        M. Suzuki, {\Journal\PRD&65&097507(2002)}.
\bibitem{suzuki}M. Suzuki, {\Journal\PRD&63&054021(2001).}
\bibitem{N_psip} X. H. Mo \etal, HEP\&NP {\bf 28}, 455 (2004).
\bibitem{BES-II} 
J. Z. Bai \etal, BES Collab., Nucl. Instr. Meth. {\bf A458}, 627 (2001).
\bibitem{J3pi}
M. Ablikim \etal, BES Collab., physics/0503001, Submitted to Nucl. Instrum. Meth. A; 
J. Z. Bai \etal, BES Collab., {\Journal\PRD&70&012005(2004)}.
\bibitem{kstark} Generator for generating $\kstark$ events
with the angular distribution of $\psip$ decays into
VP mesons. The angular distribution is described by
${\frac{d^3\sigma}{d\cos\theta_V d\cos\theta_1 d\phi_1}}$
$=\sin^2$$\theta_1[1+\cos^2\theta_V+\sin^2\theta_V\cos(2\phi_1)],$
where $\theta_V$ is the angle between the vector meson and the positron
direction. For $K^*(892)\rar K\pi$, $\theta_1$ and
$\phi_1$ are the polar and azimuth
$\pi$ with respect to the helicity direction of the $K^*(892)$.
\bibitem{kskl} J. Z. Bai \etal, BES Collab., {\Journal\PRL&92&052001(2004)};
               J. Z. Bai \etal, BES Collab., {\Journal\PRD&69&012003(2004)}.
\bibitem{BES-VT}
               J. Z. Bai \etal, BES Collab., {\Journal\PRD&69&072001(2004)}.
\bibitem{systematics} See the determination of systematic errors in
  J. Z. Bai \etal, BES Collab., {\Journal\PRD&70&012005(2004)}.
\bibitem{wangp} P. Wang, C. Z. Yuan, X. H. Mo and D. H. Zhang, {\Journal\PLB&593&89(2004)}.
\bibitem{wangp2} P. Wang, C. Z. Yuan and X. H. Mo, {\Journal\PRD&69&057502(2004)}.
\bibitem{Lum} S. P. Chi, X. H. Mo and Y. S. Zhu, HEP\&NP {\bf 28}, 1135 (2004).
\bibitem{Jconrad} J. Conrad \etal, {\Journal\PRD&67&012002(2003)}. 
We use the modified likelihood ratio ordering including the systematic 
uncertainties of signal (Gaussian parametrization) and  
background (flat parametrization) in confidence interval construction.
\bibitem{Jdecay} L. K\"{o}pke and N. Wermes, Phys. Rep. {\bf 174}, 67 (1989). 
\bibitem{Cleo}N. E. Adam \etal, {\Journal\PRL&94&012005(2005)}.
\end{thebibliography}
\end{document}